\documentclass[aps,prl,twocolumn,showpacs,superscriptaddress,groupedaddress]{revtex4}
\usepackage{graphicx} 
\usepackage{dcolumn}
\usepackage{bm}
\usepackage{amssymb}
\usepackage{amsmath}

\begin{document}

\hspace{5.2in} \mbox{}

\title{Amorphous nucleation precursor in highly nonequilibrium fluids}

\author{Gyula I.~T\'oth}
\affiliation{Research Institute for Solid State Physics and Optics, 
P. O. Box 49, H-1525 Budapest, Hungary}

\author{Tam\'as ~Pusztai} 
\affiliation{Research Institute for Solid State Physics and Optics, 
P. O. Box 49, H-1525 Budapest, Hungary}

\author{Gy\"orgy~Tegze}
\affiliation{Research Institute for Solid State Physics and Optics, 
P. O. Box 49, H-1525 Budapest, Hungary}

\author{Gergely~T\'oth}
\affiliation{Institute of Chemistry, E\"otv\"os University, P. O. Box 32, 
H-1518 Budapest, Hungary}

\author{L\'aszl\'o~Gr\'an\'asy} 
\affiliation{Research Institute for Solid State Physics and Optics, 
P. O. Box 49, H-1525 Budapest, Hungary}
\affiliation{BCAST, Brunel University, Uxbridge, Middlesex, UB8 3PH, UK}
 
\vskip 0.25cm
\date{\today}

\begin{abstract}
Dynamical density functional simulations reveal structural
aspects of crystal nucleation in undercooled liquids: the 
first appearing solid is amorphous, which promotes the nucleation of 
bcc crystals, but suppresses the appearance of the fcc and hcp 
phases. These findings are associated with features of the 
effective interaction potential deduced from the amorphous structure.

\end{abstract}

\maketitle

Mounting evidence indicates that the classical picture of crystal nucleation, 
which considers ''heterophase" fluctuations of only
the stable phase, is oversimplified.
Early analysis by 
Alexander and McTague suggests preference for bcc freezing in simple 
liquids \cite{bcc}. Atomistic simulations for the Lennard-Jones (LJ) system 
have verified that small heterophase fluctuations have the metastable (MS) bcc structure, 
and even larger clusters of the stable (S) fcc structure have a bcc interface layer 
\cite{wolde}, while the ratio of the two phases can be tuned by changing the 
pressure \cite{desgranges}. Composite bcc/fcc nuclei have also been predicted by 
continuum models \cite{cont}. Two-stage nucleation 
has been reported in systems that have a metastable critical point in the 
undercooled liquid (including solutions of globular proteins \cite{vekilov} 
and eutectic alloys \cite{toth08}); the appearance of the crystalline phase 
is assisted by liquid droplets, whose formation precedes/helps crystal 
nucleation \cite{frenkel}. Recent studies indicate a similar behavior in simple 
liquids such as the LJ \cite{lutsko} or hard-sphere (HS) \cite{schilling} 
fluids, where dense liquid/amorphous precursor assists crystal nucleation. 
Analogous behavior has been reported for colloidal systems in 2D \cite{coll2D} 
and 3D \cite{coll3D}. Brownian Dynamics studies for the HS system \cite{tanaka1_2} 
show the evolution of medium range crystalline order during the pre-nucleation 
stages. Liquid mediated crystal-amorphous and crystal-crystal transitions have 
also been predicted \cite{Levitas}. 
These findings imply that the nucleation precursors 
are fairly common. A deeper understanding of nucleation 
pathways requires a systematic study of a system, in which amorphous and crystalline 
structures compete during solidification.

Such a system is defined by the single-mode phase-field crystal (1M-PFC) model of 
Elder {\it et al.} \cite{elder}, a simple dynamical density functional theory, 
which has bcc, fcc, and hcp stability domains \cite{phdiag}, and the appearance 
of amorphous 
phase and two-step nucleation has also been reported \cite{berry}. A two-mode 
extension of the model by Wu {\it et al.} (2M-PFC) has been designed to promote 
fcc solidification \cite{wu_fcc}, whereas with a specific choice of model parameters 
the 1M-PFC model can also be recovered. 

Herein, we address crystal nucleation in PFC models interpolating
between the 1M-PFC and 2M-PFC limits. 

\begin{figure}
\includegraphics[width=1\linewidth]{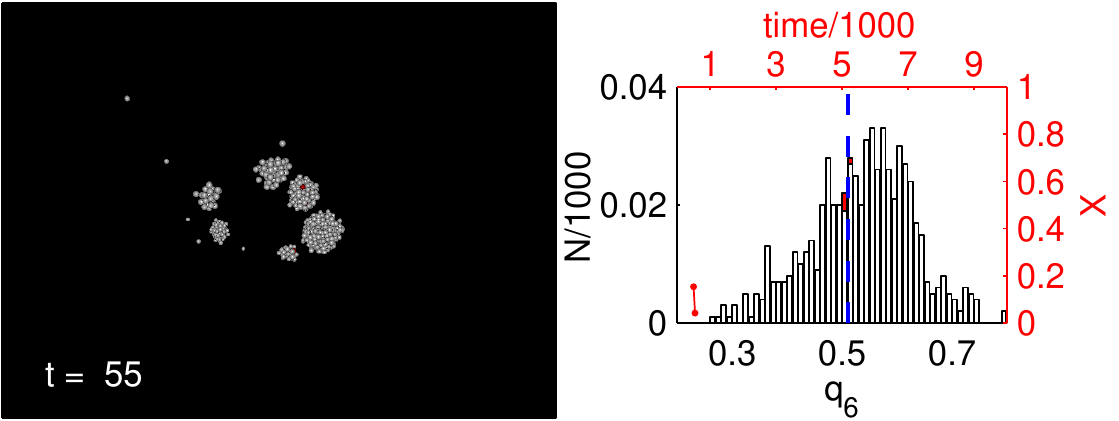}
\includegraphics[width=1\linewidth]{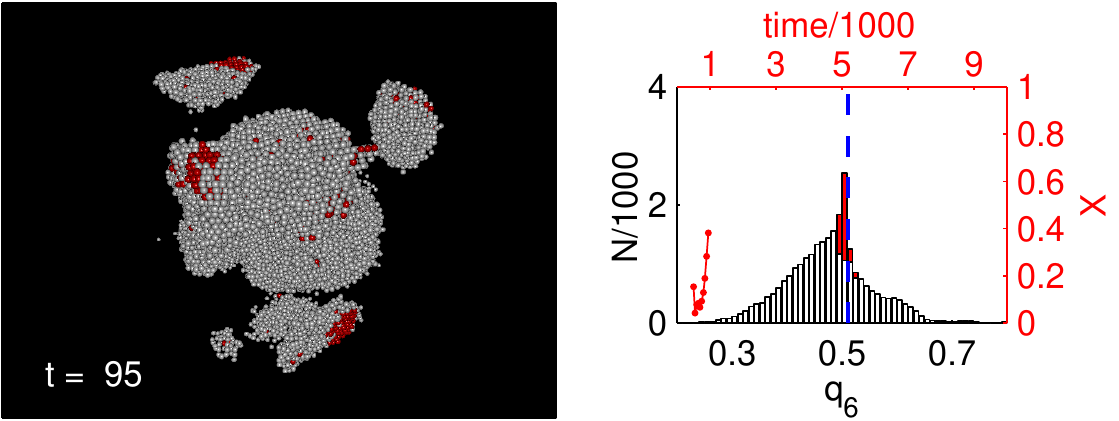}
\includegraphics[width=1\linewidth]{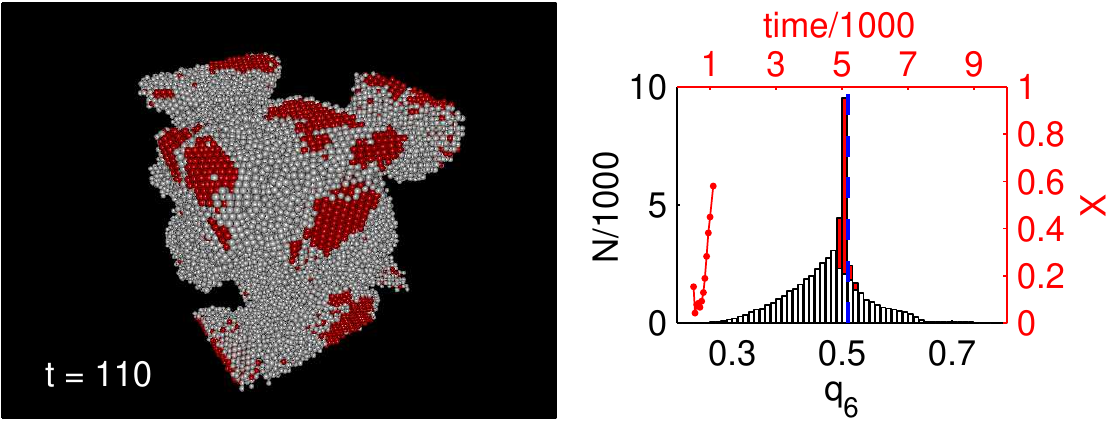}
\includegraphics[width=1\linewidth]{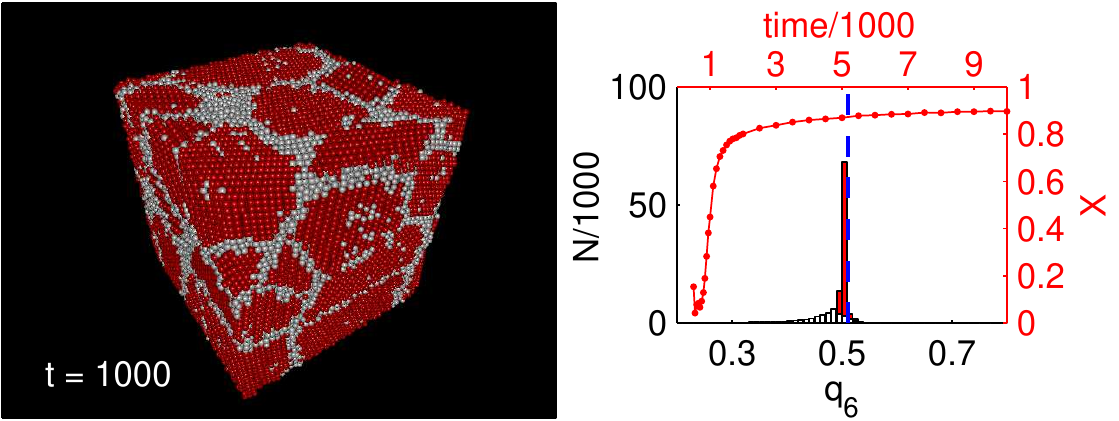}
\caption{(color online) Two-step nucleation in the 1M-PFC model at 
$\epsilon = -0.1667$ and 
$\psi_0 = -0.25$. 
Left: Snapshots of the density distribution taken at the dimensionless times 
$\tau = 57.74 t$. Spheres of diameter of the interparticle distance 
centered on density peaks higher than a threshold (0.15) are shown 
that are colored red if $q_4 \in [0.02, 0.07]$ and 
$q_6 \in [0.48, 0.52]$ (bcc-like), and white otherwise. 
Right: population distribution of $q_6$ (histogram painted 
similarly), and the time-dependence of the fraction $X$ of 
bcc-like neighborhoods (solid line). 
Note the nucleation of amorphous clusters and the formation
of amorphous grain boundaries \cite{aGB}.}
\label{nucl}
\end{figure}

\begin{figure}
\includegraphics[width=1\linewidth]{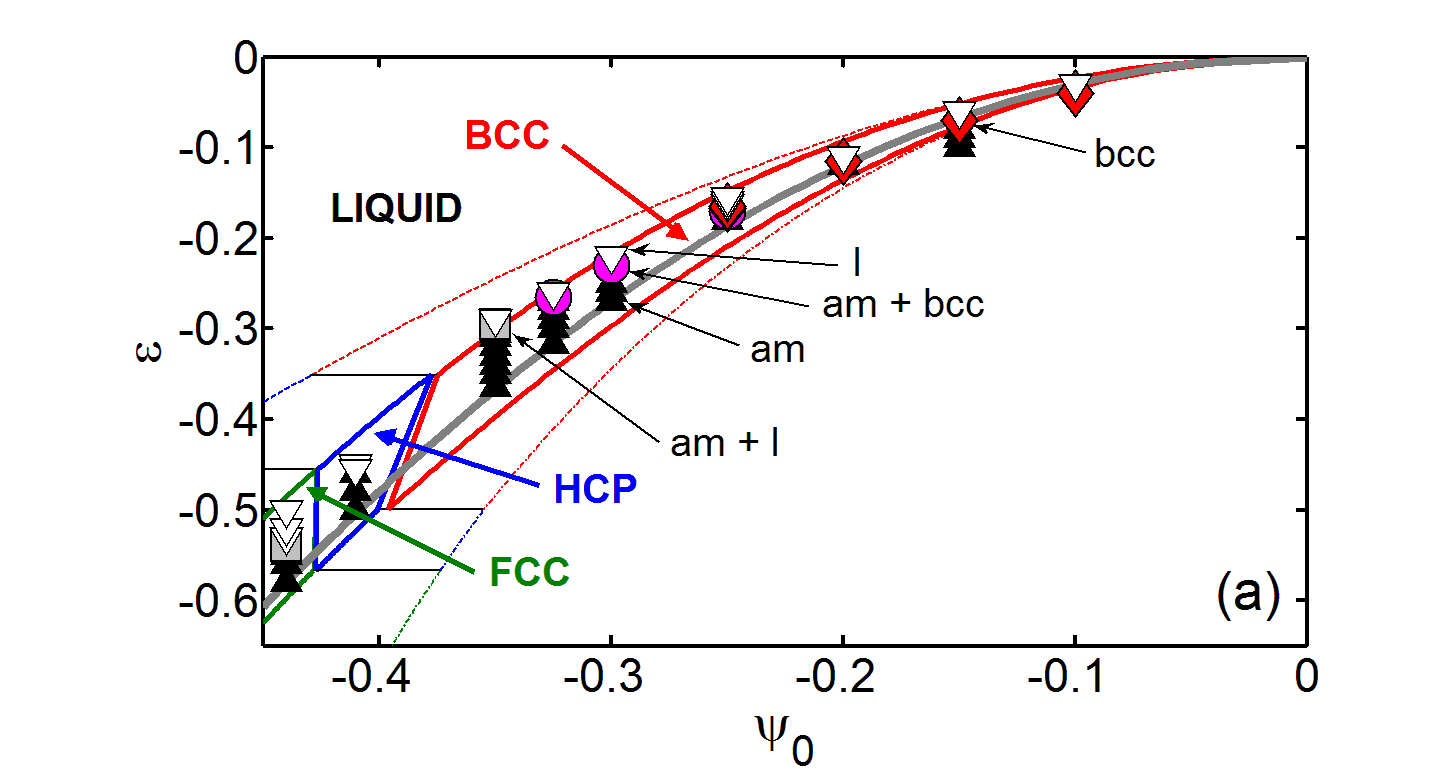}
\includegraphics[width=1\linewidth]{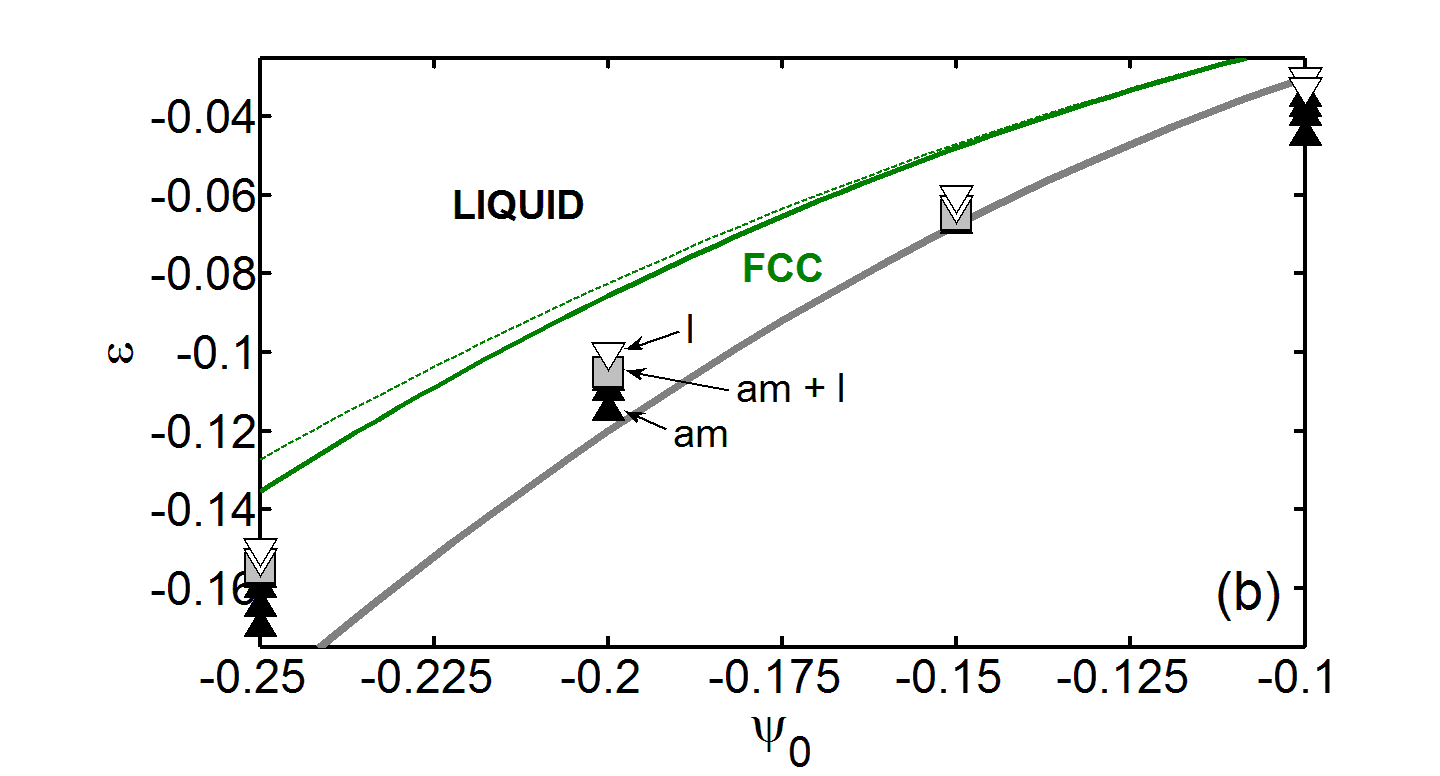}
\includegraphics[width=1\linewidth]{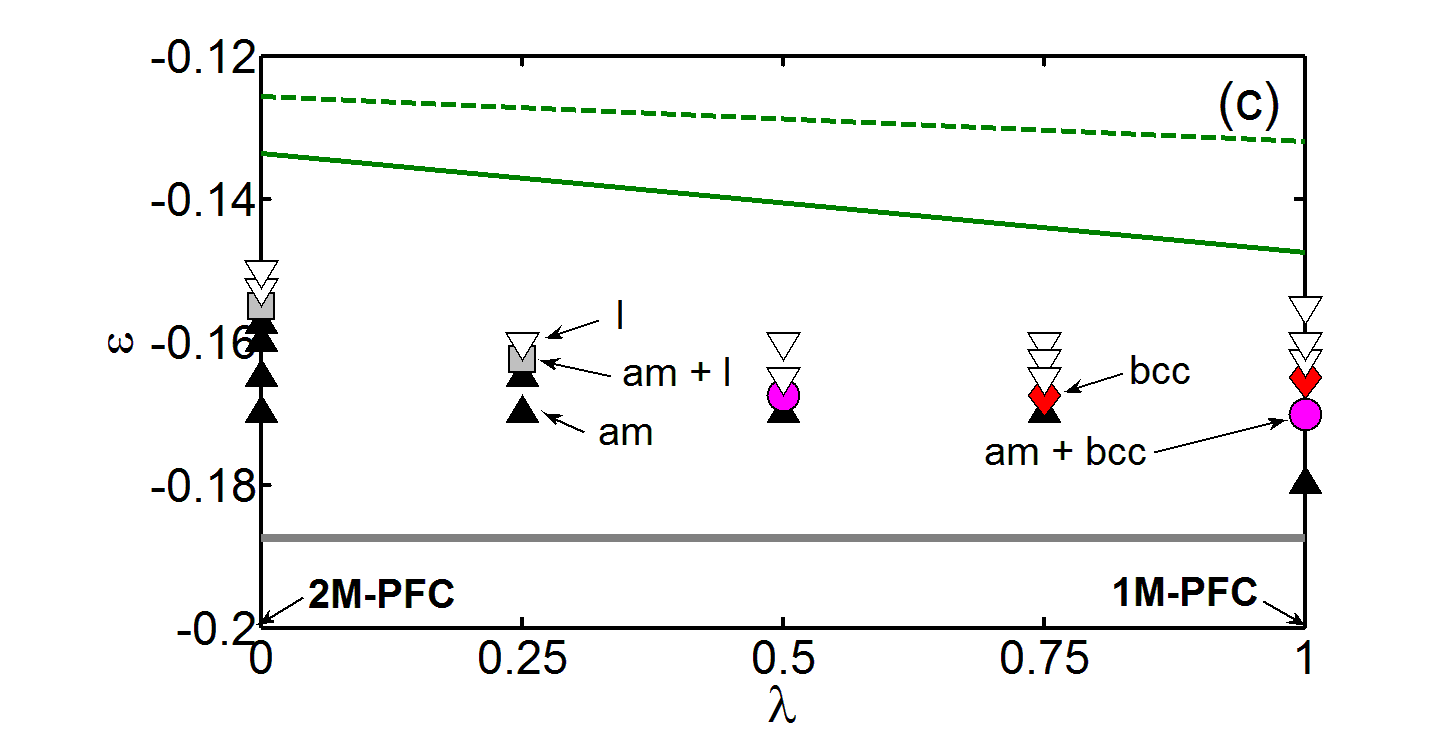}
\caption{(color online) Nucleation map for PFC models: (a) 1M-PFC; (b) 2M-PFC; 
(c) dependence on $\lambda$ at $\psi_0=-0.25$. 
The phase content obtained after $10^5$ time steps is shown: open triangle 
-- liquid; square -- (amorphous + liquid); circle -- (amorphous + bcc); 
diamond -- bcc; full triangle -- amorphous. The heavy grey line stands for 
the stability limit of the liquid. Parts of the phase diagram are also shown.}
\label{map}
\end{figure}

\begin{figure}
\includegraphics[width=1\linewidth]{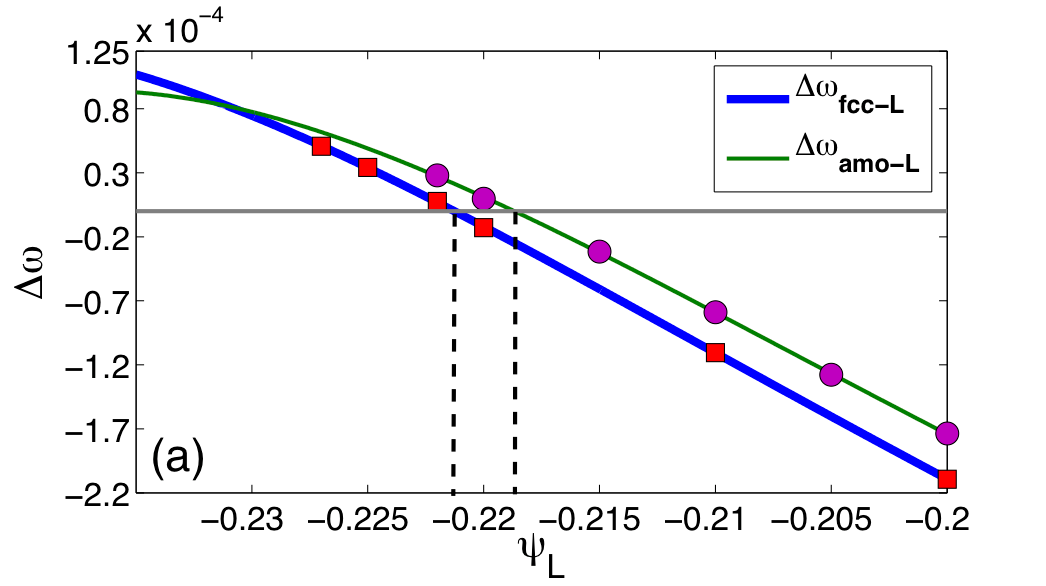}
\includegraphics[width=1\linewidth]{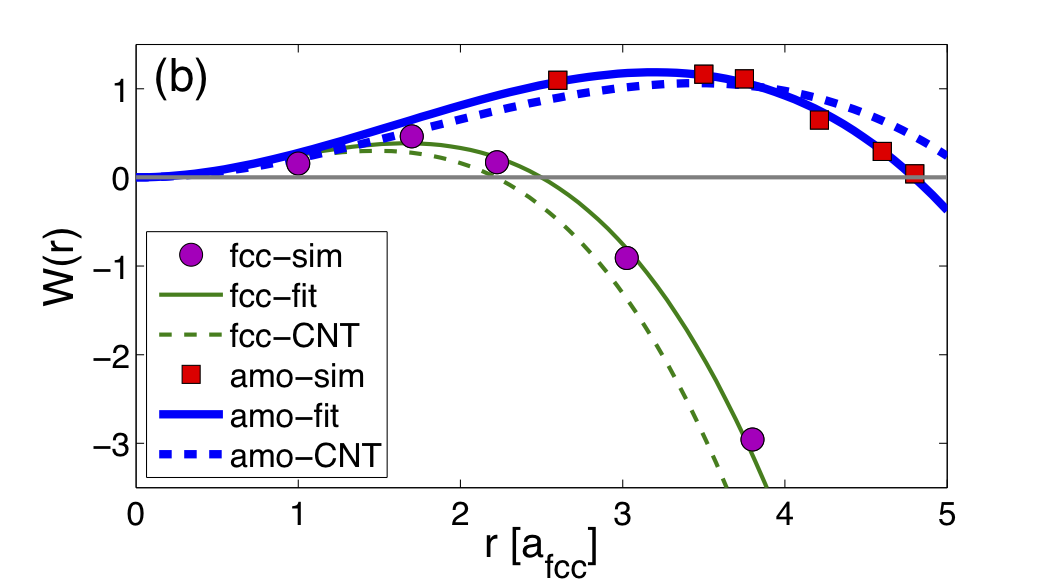}
\caption{(color online) Properties of the amorphous phase in the 2M-PFC limit 
from the ELE at $\epsilon = -0.1$: (a) Driving force for the liquid to amorphous
transition ($\Delta \omega_{amo-L}$). For comparison the driving force for 
fcc freezing ($\Delta \omega_{fcc-L}$) is also shown. 
The vertical lines indicate liquid densities at the
(S) fcc-liquid and (MS) amorphous-liquid coexistences.
(b) Nucleation barrier for the amorphous and fcc phases at $\psi_0 = -0.21775$. 
($a_{fcc}$ -- lattice constant of the fcc structure.) 
Fits of $W = Ar^3+Br^2$ (solid lines) are compared with 
estimates from the classical droplet model using the equilibrium 
interface free energy and the driving force (dashed lines).}
\label{ELE}
\end{figure}

First, we re-cast the free energy of the PFC models in terms 
of $\lambda = R_1/(1+R_1) \in [0,1]$ ($R_1$ is the 
relative strength of the first- and second-mode contributions \cite{wu_fcc}), 
a parameter that can be used to interpolate between the 2M-PFC ($\lambda = 0$) 
and 1M-PFC ($\lambda = 1$) models:
\begin{equation}
\label{F_functional}
\begin{split}
   &\mathcal{F} = \int d\mathbf{r} \left\{\frac{\psi}{2}\left[\epsilon+(1+\nabla^2)^2 \right.\right. \\
   & \left. \left. \times (\lambda+\{1-\lambda\}\{Q_1^2 + \nabla^2\}^2)\right]\psi + \frac{\psi^4}{4}\right\},
\end{split}
\end{equation}
\noindent
where $\psi \propto (\rho - \rho_L^{ref})/\rho_L^{ref}$ is the scaled density 
difference relative to the reference liquid of particle density $\rho_L^{ref}$. 
The reduced temperature $\epsilon$ can be related to the bulk moduli 
of the fluid and the crystal, whereas $Q_1 = q_1/q_0$ $(= 2/3^{1/2}$ for fcc \cite{wu_fcc}) 
is the ratio of the wave numbers corresponding to the two modes. 

The respective dimensionless Euler-Lagrange equation (ELE) and 
equation of motion (EOM) read as 
$   \frac{\delta \mathcal{F}}{\delta \psi} = 
   \left(\frac{\delta \mathcal{F}}{\delta \psi}\right)_{\psi_0},$
and
$\frac{\partial \psi}{\partial \tau} = 
   \nabla^2\frac{\delta \mathcal{F}}{\delta \psi}+\zeta,$
respectively,
where $\frac{\delta \mathcal{F}}{\delta \psi}$ denotes the functional derivative 
of $\mathcal{F}$ with respect to $\psi$, and $\tau$ is the dimensionless time. 
The RHS of the ELE is taken at the far-field value $\psi_0$ (homogeneous liquid). 
In the EOM, the fluctuations are represented by a colored Gaussian noise $\zeta$ of 
correlator $\langle\zeta(\mathbf{r},\tau)\zeta(\mathbf{r}',\tau')\rangle = 
-\alpha \nabla^2g (|\mathbf{r} - \mathbf{r}'|,\sigma)  \delta (\tau - \tau')$,
where $\alpha$ is the noise strength, and $g (|\mathbf{r} - \mathbf{r}'|,\sigma)$ 
a high frequency cutoff function \cite{ojalvo_book} for wavelengths shorter than the 
interatomic spacing ($\sigma$). Due to the overdamped conservative dynamics the 
EOM realizes, the PFC models defined so are suitable for describing crystalline colloidal 
aggregation \cite{EOM_lowen,pfc_coll}. These equations have been solved 
numerically \cite{jcompp} on rectangular grids of typical size of
$512 \times 256 \times 256$ (ELE) and $256 \times 256 \times 256$ (EOM), 
assuming a periodic boundary condition. 
The ELE has been used to determine the phase 
diagram, the driving force, the nucleation barrier, 
and the coexistence properties, including the densities and the
solid-liquid interface free energy (as described in \cite{phdiag}), 
while the EOM has been applied to simulate nucleation. Owing to the effect of noise
on the free energy, 
the results from the two approaches 
converge for $\zeta \rightarrow 0$. 

The results of the nucleation studies performed solving the EOM under 
condition described in \cite{conditions} are summarized in Figs. \ref{nucl} and 
\ref{map}. In the case of the 1M-PFC model, in a large part of the bcc stability 
domain we have observed two-step nucleation starting with formation of amorphous 
clusters, in which the bcc phase nucleates subsequently (see Fig. \ref{nucl}). We 
have used the $q_4$ and 
$q_6$ order parameter to characterize the local structure \cite{wolde}.
With increasing undercooling, the nucleation rate of the amorphous clusters increases,
leading to spatially nearly homogeneous transition at high undercoolings. In contrast, 
we have not detected any phase transition for more than $10^6$ time steps at 
$\epsilon = -0.1598$. These findings strongly indicate that crystal nucleation 
is enhanced by the amorphous precursor, and that direct bcc crystal nucleation
from the liquid requires orders of magnitude longer time than via the precursor. 
This behavior appears analogous to the role the non-crystalline precursor plays in 
colloids \cite{coll2D,coll3D} and simple liquids \cite{lutsko, schilling}.

We were unable to nucleate crystalline phases other than bcc in the 1M-PFC model 
[Fig. \ref{map}(a)]. Even in the stability domain of the hcp and fcc phases, the 
amorphous phase formed in the 
time window of the simulations. Remarkably, this stayed so even in the 2M-PFC limit 
[Fig. \ref{map}(b)]. Interestingly, the amorphous phase appears
to coexist with the liquid, indicating a first-order phase transition between 
these phase, in agreement with the observed nucleation of the amorphous 
state. This suggests significant differences (e.g., in density) between 
the liquid and amorphous phases. Varying $\lambda$ at $\epsilon = -0.1$, we see 
a gradual transition from the 1M-PFC behavior (liquid $\rightarrow$ amorphous 
$\rightarrow$ bcc) to the behavior seen on the 2M-PFC side (liquid $\rightarrow$ 
amorphous) [Fig. \ref{map}(c)]. (Comparable results were obtained for constant 
cooling rates.)

To investigate the lack of fcc crystallization in the 2M-PFC model
specifically designed to crystallize to the fcc phase, we used the ELE 
at $\epsilon = -0.1$ for determining the free energy $\gamma$ of the 
fcc-liquid and amorphous-liquid interfaces (Table I), and the driving 
force $\Delta \omega$ for fcc freezing and amorphization [Fig. \ref{ELE}(a)] 
(see the methodology in \cite{phdiag}), whose interplay determines the 
nucleation rate \cite{prefactors}. 
In the density range of interest the fcc phase is preferred thermodynamically
[Fig. \ref{ELE}(a)].
Between the fcc-liquid and amorphous-liquid coexistences, there is no driving 
force for amorphization, 
so fcc freezing should take place. To evaluate $\gamma$, we have created equilibrium 
sandwiches (fcc-liquid-fcc and amorphous-liquid-amorphous), solved the 
ELE, and determined the grand potential emerging from the two interfaces. 
Comparable density changes were found at the amorphous-liquid and fcc-liquid 
transitions (Table I), a finding consistent with the first order amorphization 
transition implied earlier. Remarkably, 
$\gamma_{am-L}^{eq} \approx 0.67 \gamma_{fcc-L}^{eq}$ (Table I). 
The nucleation barriers calculated for $\psi_0 = -0.21775$ using the classical 
droplet model, $W(r) = (4\pi/3) r^3 \Delta \omega + 4\pi r^2 \gamma$,
are shown in [Fig. \ref{ELE}(b)]. At this supersaturation fcc nucleation 
is clearly preferable. Yet, we have not seen freezing even after $1.9\times10^7$ 
time steps. The liquid density beyond which $W_{fcc} > W_{am}$ is 
$\psi_0 \approx -0.21$. This is consistent with 
the finding that in the density range, where solidification 
could be observed at all ($\psi_0 \geq -0.1962$ for $10^7$ time steps), 
the amorphous phase nucleated. It appears that we cannot observe fcc nucleation 
because of a technical difficulty: the time accessible for simulations 
is too short. Note that the dynamic EOM studies and the equilibrium ELE results 
consistently indicate separate time scales for changes of density and structure.

\begin{table}[b]
\caption{Equilibrium densities ($\psi^{eq}$) and interface free energies ($\gamma$) in
equilibrium (eq) and from fitting to $W(r)$ (fit) 
for the 2M-PFC model at $\epsilon = -0.1$.}
\label{table:ifvelo}
\begin{ruledtabular}
\begin{tabular}{cccccccc}
X & $\psi_L^{eq}$ & $\psi_S^{eq}$ & $\gamma_{X-L}^{eq}$ & $\gamma_{X-L}^{fit}$ \\
\hline
am & $-0.21885$ & $-0.21404$ & $1.79\times10^{-4}$ & $2.34\times10^{-4}$\\
fcc & $-0.22139$ & $-0.21629$ & $2.76\times10^{-4}$ & $2.80\times10^{-4}$\\
\end{tabular}
\end{ruledtabular}
Subscripts S and L stand for solid and liquid, respectively.
\end{table}

\begin{figure}
\includegraphics[width=1\linewidth]{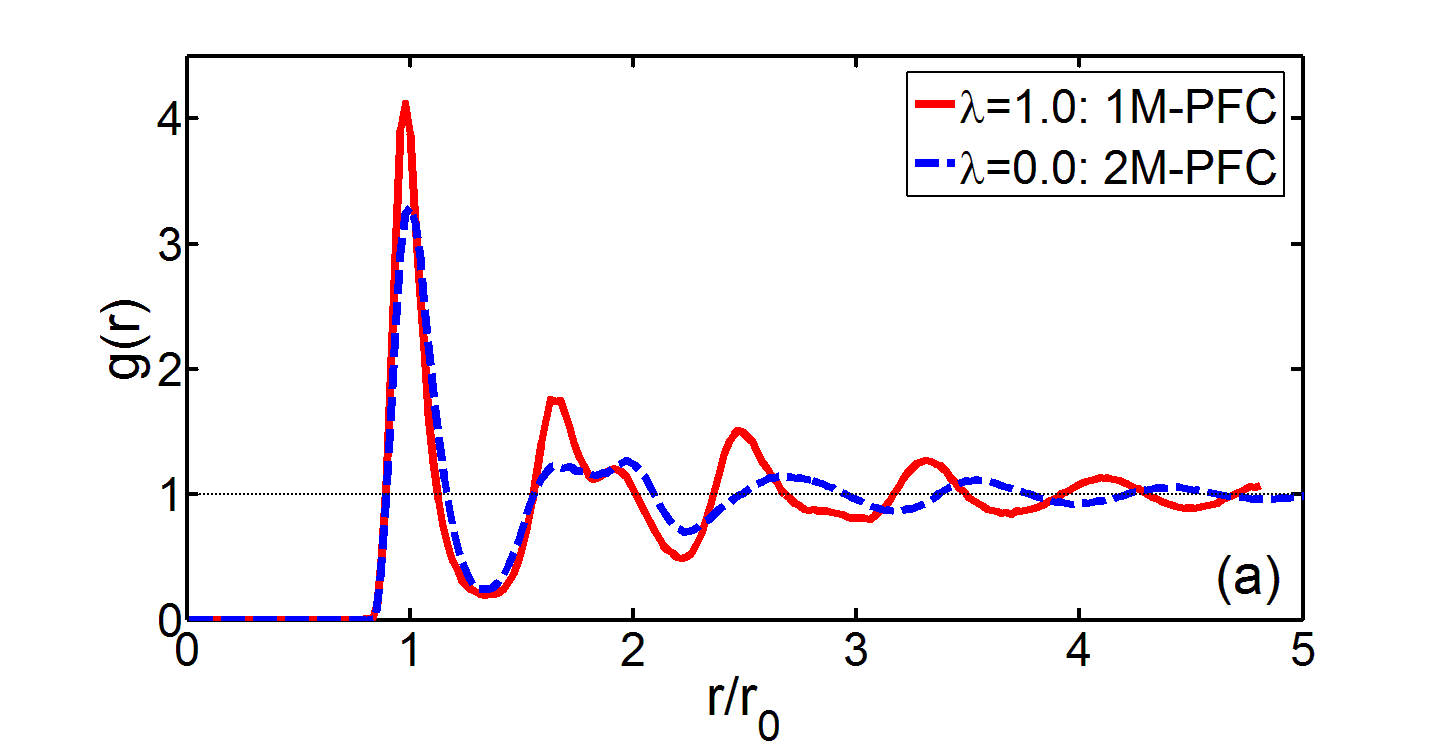}
\includegraphics[width=1\linewidth]{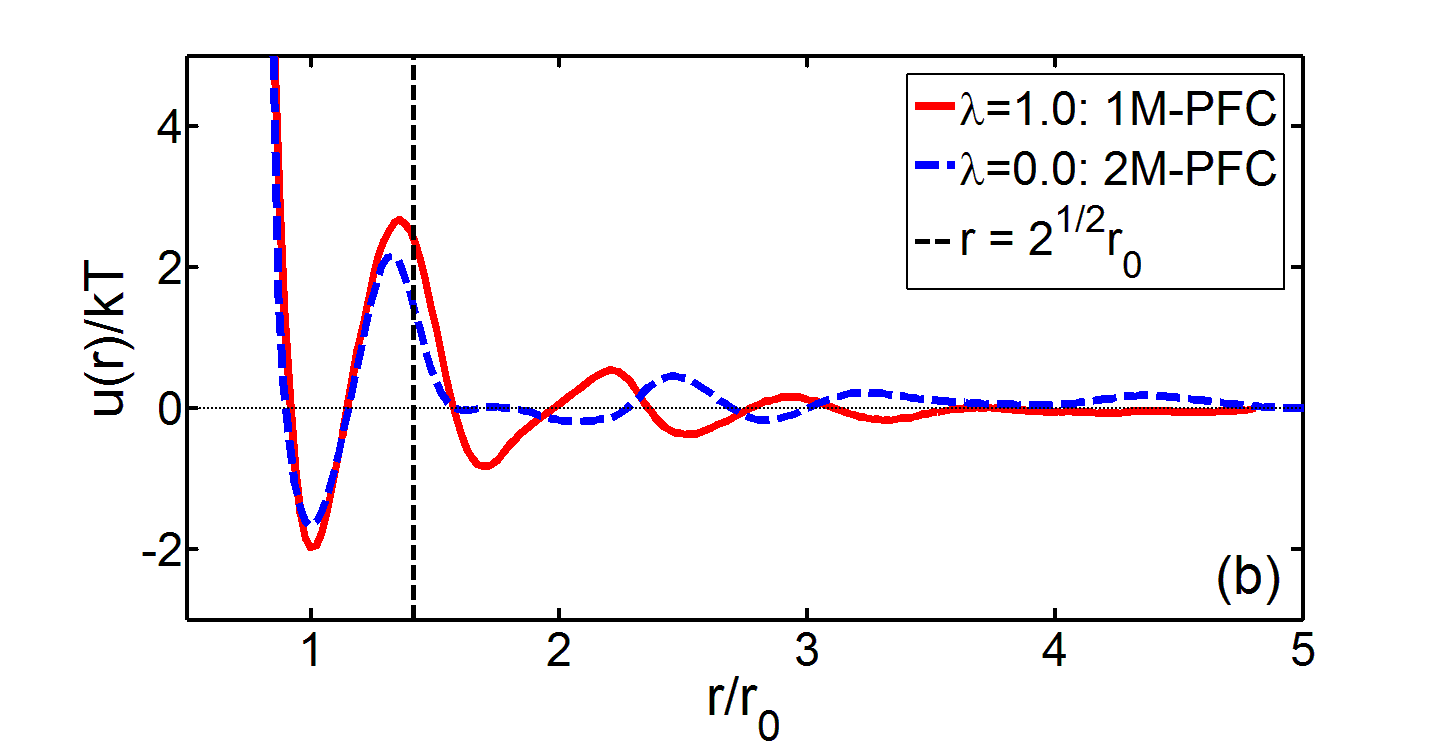}
\caption{(color online) Amorphous structure and effective pair potential in 
the 1M- and 2M-PFC models: 
(a) $g(r)$ of the amorphous structures. (b) Effective pair potentials evaluated from
$g(r)$ by Schommers' method.}
\label{struct}
\end{figure}

For the PFC models the free energy barrier has a rough surface with
many local minima that can be mapped out directly by using the ELE \cite{phdiag}. 
This provides an independent test of the previous computations. The results shown 
in [Fig. \ref{ELE}(c)] indicate a reasonable agreement between
the predicted (dashed lines) and the directly evaluated barriers (symbols + solid 
lines). This is reflected in the similarity of the interfacial properties evaluated 
in equilibrium and from fitting the droplet model (see Table I). The minor discrepancy
presumably originates from the fact that for such nanoclusters the classical 
droplet model is probably not very accurate. 

To rationalize the dominance of amorphous solidification
in a substantial part of the phase diagram, we have evaluated 
effective pair potentials for the 1M- and 2M-PFC models from the pair 
correlation function of the respective amorphous phases [Fig. \ref{struct}(a)] 
using Schommers' iterative method \cite{schom} that works reasonably for single 
component systems \cite{toth}. The potentials obtained are similar for short 
distances, and have a peak at $\sim r_0 \sqrt{2}$, where $r_0$ is the radius 
at the main minimum of the potential [Fig. \ref{struct}(b)]. 
Remarkably, such potentials have been designed to realize monatomic 
glassformers, as a peak at $r_0 \sqrt{2}$ suppresses the close 
packed crystal structures \cite{doye}. Hence, we associate the evident 
difficulty to produce the fcc and hcp phases with this feature of the PFC 
effective interaction potentials. Furthermore, in the presence 
of multiple minima of the interaction potential coexistence of disordered 
phases is expected \cite{polyamorphism}, as indeed seen here.    

Summarizing, the PFC models display MS amorphous-liquid coexistence 
and first-order amorphization. In the cases accessible for dynamic 
simulations, the nucleation of the amorphous phase is faster than 
crystal nucleation. This leads to a separation of time scales for density 
and structural changes, as seen in other systems \cite{schilling}. 
However, some details differ: Such coexistence is unknown in the HS system,
while the fcc and hcp structures are suppressed here. It is also unclear whether 
along the reaction coordinate specified in Ref. 8, the free energy landscape 
of the PFC models is similar to that of the LJ system. 

Combining the results obtained for various potentials (LJ, HS, the present PFC 
potentials, etc. \cite{lutsko,schilling,doye,polyamorphism}), 
it appears that a repulsive core suffices for the 
appearance of a disordered precursor, whereas the peak at $\sim r_0 \sqrt{2}$ 
correlates with the observed suppression of fcc and hcp structures, while the 
coexistence of the liquid and amorphous phases seen here can be associated 
with multiple minima of the interaction potential.

\begin{acknowledgments} This work was supported by the EU FP7 
Project ENSEMBLE under Grant Agreement NMP4-SL-2008-213669
and by TAMOP 4.2.1B-09/1/KMR-2010-0003.
\end{acknowledgments}


\begin{thebibliography}{99}
\bibitem{conditions} $\Delta x = 1.0; \Delta \tau = 0.5774; \alpha = 0.42;$ 
the number of time steps was $10^5$, if not stated otherwise. The simulations started 
with a homogeneous liquid of density $\psi_0$.
\bibitem{prefactors} The nucleation prefactors are expected to be 
similar for crystallization and amorphization: both are 
determined by the molecular mobility in the liquid state.
\end{thebibliography}
\end{document}